\documentclass[twocolumn,superscriptaddress]{revtex4}

\usepackage{amsfonts}
\usepackage{epsf,psfig}
\usepackage{graphics}

\begin{document}
\frenchspacing
\title{A fitness model for the Italian Interbank Money Market}
\author{G. De Masi}
\affiliation{Dipartimento di Fisica, Universit\`a di L'Aquila, Via Vetoio, 67010 
Coppito (AQ), ITALY and  
Dipartimento di Fisica, Universit\`a di Roma ''La Sapienza'', P.le A. Moro 5, 
00185 Roma, ITALY}
\author{G. Iori}
\affiliation{Department of Economics, City University, Northampton  
Square, EC1V 0HB London, UK}
\author{G. Caldarelli}
\affiliation{INFM-CNR Centro SMC and Dipartimento di Fisica 
Universit\'a di Roma "La Sapienza" Piazzale Moro 5, 00185 Roma, Italy, 
and Centro Studi e Museo della Fisica Enrico Fermi, Compendio Viminale, 
00185 Roma, Italy}

\begin{abstract}
We use the theory of complex networks in order to 
quantitatively characterize the formation of communities in a particular 
financial market. The system is composed by different banks exchanging on a 
daily basis loans and debts of liquidity. 
Through topological analysis and by means of a model of network growth we 
can determine the formation of different group of banks characterized by 
different business strategy. 
The model based on Pareto's Law makes no use of growth or preferential 
attachment and it reproduces correctly all the various statistical 
properties of the system.
We believe that this network modeling of the market 
could be an efficient way to evaluate the impact of different 
policies in the market of liquidity.
\end{abstract}
\maketitle

Co-evolution and interaction between different agents is known to be one of the 
ingredients of the so-called 
complex systems. Several examples can be found in social\cite{CFSV06,CSCBDLC06}, 
biological\cite{GCP03,MMC00,VD05,H05}, economical\cite{CMZ97} and technological systems\cite{K99}. 
Any of these systems is composed by a set of agents competing and sometimes receiving 
reciprocal advantage interacting each other. 
In the above situation both coalition and competition are at the basis of the process of 
co-evolution and self-organization of the system.
While this class of problems has been traditionally studied in game theory, more recently it has 
been introduced an approach based on graph theory\cite{BCLM03,OCKKK03}
By using networks\cite{AB02,N03},  we can characterize quantitatively the 
interaction between agents by means of a series of topological quantities. 
The case of study presented here is composed by banks operating in the Italian 
market\cite{IDPGC06}. Banks try to maximize their returns given some constraints from the 
European Central Bank. This complex interaction results in a differentiation of the strategies
that is well described by means of graph cliques.
More specifically banks of the same size tend to form a cluster and to adopt a 
similar business strategy. 

A network is a mathematical object composed by vertices and edges joining them. 
Different measures can be made, 
from the degree distribution (the degree is the number of edges per vertex) to the diameter 
(i.e. the maximum of the distances between every couple of vertices). 
It is interesting to note that 
different real world networks (ranging from social to biological ones), display a 
a scale-free distribution of  degrees and a ``small-world'' character, that is to say the 
diameter is usually very small\cite{S01}. 
More complicated measures determine also the presence of communities 
in a network. In this case, some methods have been proposed\cite{GN02,CSCC05,RB04} 
but no general approach is available.  
%The determination of communities is very important for our purposes, since 
%it determines the set of agents with similar behavior. By detecting common behaviors of 
%communities of vertices would help in providing insights to understand the of hidden 
%organization principles. Network communities might indeed represent in a simplified 
%way the emergent dynamic of the system.
\begin{figure}
\psfig{figure=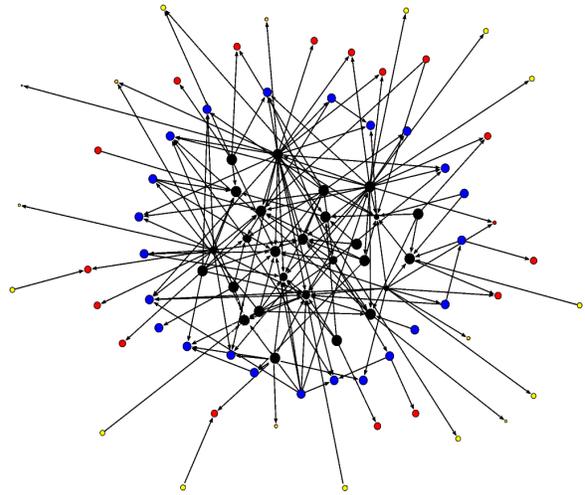,height=6.5cm}  
\caption{(Color on line) A plot of the inter bank network. The color codes for the various 
groups are the following: $1$=yellow, $2$=red, $3$=blue, $4$=black. 
Note that the dark vertices (bank of group $4$) form the core of the system.}
\label{f1}
\end{figure} 
The set of banks with their internal loans and debts has a structure than can be 
naturally described by means of a network. In this case 
the vertices are the different banks. For every  pair of banks 
$i$ and $j$ we draw an oriented edge from $i$ to $j$, if bank $j$ borrows liquidity from bank $i$. 
The number of in-coming and out-going edges of a vertex is called respectively the 
in-degree $k_{in}$ and the out-degree $k_{out}$ of the vertex (their sum gives the total 
degree $k$). The loans are originated by the fact that every bank needs liquidity 
in order to satisfy demands of customers. To buffer liquidity shocks the 
European Central Bank requires that on average $2\%$ of all deposits and debts owned 
by banks are stored in national central banks. Given this constraint, 
banks can exchange excess reserves on the interbank market with the objective to 
satisfy the reserve requirement and in order to minimize the reserve implicit 
costs\cite{HMM01,BIR04,BES04}. 
The data set analyzed is the  e-MID(ref dataset)\cite{EMID}. This data set is 
composed by $586,007$ overnight 
transactions (i.e. payments of loans must be done in $24$ hours) concluded from 
January $1^{st} 1999$ to December $31^{st} 2002$ . The network is composed by a set of $N$ banks (the average number of $\langle N \rangle$ banks daily active is $140$) connected by an average number of links $\langle L \rangle = 200$ (in case of multiple transactions among banks $i$ and $j$, we count just one link). As in many other complex networks we find here a 
fat tail distribution. By fitting these data with a power law we obtain for the 
total degree a frequency distribution 
$F(k)\propto k^{-2.3}$ and a similar behavior for the in/out-degree with 
exponents $F(k_{in})\propto k_{in}^{-2.7}$and 
$F(k_{out})\propto k_{out}^{-2.15}$. 
Regardless the precise form of the fit, the fat-tail indicates 
that banks have an highly heterogeneous 
behavior, since the number of their partners varies very widely.  
We also measure the assortativity and the clustering coefficient of the network. 
The first one is defined as the average value $k_{nn}(k)$ of the neighbors of a 
vertex whose degree is $k$. We find $k_{nn}(k)\propto k^{-0.5}$. 
This means that banks with few partners interact with banks with many partners. 
Conversely (on average) banks with many partners interact with banks with few or one. 
The clustering coefficient instead accounts for the number of triangles a vertex of 
degree $k$ belongs to. Also this quantity has a power law behavior of the 
kind $c(k)\propto k^{-0.8}$.  
All these measurements refer to daily networks resulting from 
composing all transactions of every day.
In fact, the system is characterized by a typical time scale of the system, 
the month. This time-scale arises from the above mentioned requirement 
from European Central Bank. The $2\%$ to be deposited in national central banks are 
computed every month (the $23^{rd}$). 
The day in which this happens (also indicated as End of Month or EOM) 
witness a frantic activity of the banks. 
Interestingly, regardless the change in volumes all the above topological 
measurements remain similar when computed in different days of month.

We try to understand if there are some banks with similar behavior and if they have 
some properties in common. We have been able to identify specific features for banks 
of different capital size. In fact for each bank we know only its category (small, medium, 
large, very large) based on the capital of the banks (as recorded by Bank of Italy). 
Nevertheless we observe that this classification is strongly correlated with the total amount of 
daily volume of transactions: we use this latter quantity as it is strictly related to capital size. Using this quantity we can divide banks in 
four groups (same number of classes of the Bank of Italy classification). 
{\em Group 1} with volume in the range $0-23$ million Euro per day, 
{\em Group 2} in the range $23-70$ million Euro per day, 
{\em Group 3}  in the range $70-165$ million Euro per day, {\em Group 4} 
over $165$ million Euro per day. 
In this way we find an overlap of more than $90\%$ between the two classifications. 

\begin{figure}
\psfig{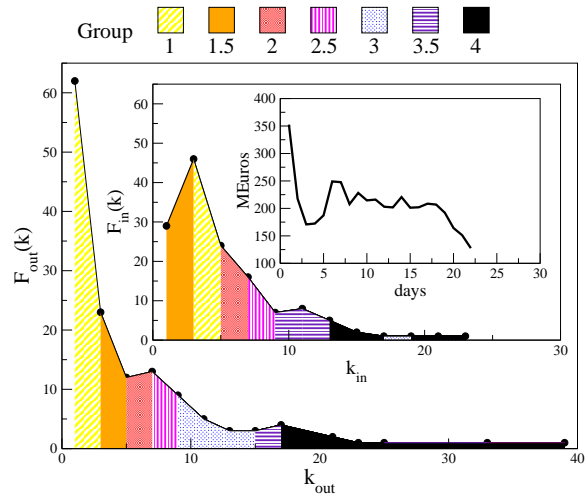}  
\caption{(A plot of the out-degree and in-degree (in the first inset) distributions 
respectively. As already noticed, the contribution to the tail of frequency 
distribution emerges from the banks of group $4$. Using the division in $4$ groups, 
i.e. in $4$ colors, mentioned in the text, we also colored each bin of 
$F(k)$ with the average color of vertices which are in that bin. 
For example the average color of banks with 
degree $10$ is blue. For non integer value of this average we introduced 
intermediate colors. 
In the smaller inset the daily volume of transactions during one month period.
}
\label{f2}
\end{figure}

Using this information we realized a picture of the system as an oriented network 
whose size and color of the vertices represent the different groups that play 
the role of communities when described by means of a network. As evident from 
Figure \ref{f1} we find that the core of the structure is composed by banks of the last 
groups (very large). The edges in Figure \ref{f1} represent the net amount of money exchanged 
in a whole day. As mentioned above the measurements in different days give similar results. 
A more quantitative measure of the different behavior of banks from different groups is 
given in the Table \ref{table_cluster}, where for every pair of groups we reported the mean 
percentage of the 
total number of transactions between banks of those groups. 
This result is confirmed by 
the first two plots of Figure \ref{f2}, where we represented in-degree frequency distribution 
(number of  borrowing edges) and the out-degree frequency distribution 
(number of lending edges) in the network (experimental distributions are obtained on an ensemble of daily networks). It is possible to compute the group of the banks 
whose degree is $k$. We represented this information by coloring accordingly the plot. 
We have separately informations about degree and volumes of different banks. 
Interestingly we note that the degree and the volume are correlated\cite{BBV02}, 
since $v(k)\sim k^{1.1}$. 

With respect to the scale of 
colors in Figure \ref{f1}, we also added some intermediate colors to account for the 
values between one group and another. The tail of the two distributions is black, 
i.e. it is mainly composed by banks of group $4$.  We again find that banks of groups 
1 and 2 are the leaves of the network, staying at periphery of the structure and 
not interacting each other. This particularity together with the experimental evidence 
that they are more lenders on average means that banks of these groups are the 
lenders for the whole system. 

\begin{figure}
\psfig{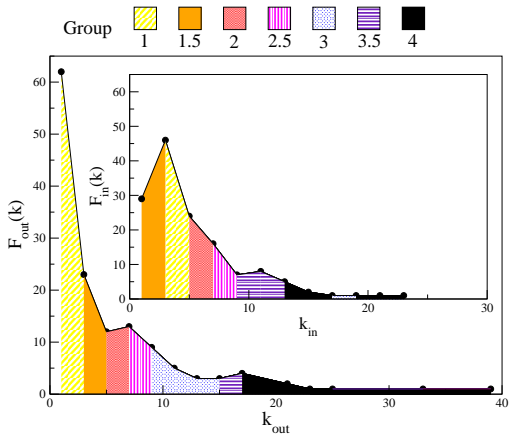}  
\caption{Left: frequency distribution $F(k)$ for a 
certain degree $k$. Comparison between experiment 
(red dots) and results obtained with simulation of our model (black dots).
Right: Above comparison between experiment (red dots) and results obtained with 
simulation of our model (black dots) for the assortativity $<knn(k)>$ and below
for the clustering coefficient $c(k)$.}
\label{f2bis}
\end{figure} 

\begin{figure}
\centerline{
\psfig{figure=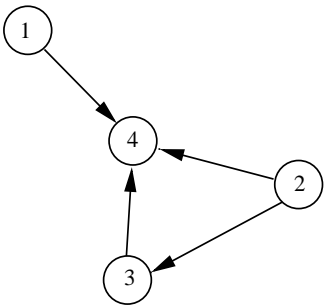,height=4cm}  
}
\caption{The division on classes of vertices permits to represent in a very easy 
way the organizational principles of the network. Following results of 
Table 1 we draw a link among two groups when the number of 
links between banks belonging to them is bigger then the average value ($10$). 
Using the net volumes as 
weight of links, we can represent the directed interactions among classes of nodes: 
class $4$ appear to be clearly a borrower and class $1$ lender.}
\label{f3}
\end{figure}

The role of the different groups is shown in the Figure \ref{f3}. 
Another measure of the clustering of banks in different groups is given by 
the volume-volume correlation $v_{nn}(v)$, that is the average value $v_{nn}$ of the 
neighbors of a vertex whose volume is $v$, 
In fact we find that $v_{nn}(v)$ is the superposition of a power-law function 
$v_{nn}(v)\sim v^{-0.3}$ with a function peaked around volume values of banks of group 1. 

In order to reproduce the topological properties we define a model whose only assumption 
is that a vertex is solely determined by its size (as measured by its capital or equivalently 
by its group). Therefore, the idea is that the vertices representing the banks are defined 
by means of an intrinsic character corresponding to the size of the bank\cite{CCDM02,SBC04}. 
Since 
this information is not available we use the total daily volume of transactions as a 
good measure of the size of banks (we stated above that this is a good approximation). We call this quantity  {\em fitness} of the bank; this is the main quantity driving the network formation in the our model.

\begin{figure} [htpb]
\centerline{
\psfig{figure=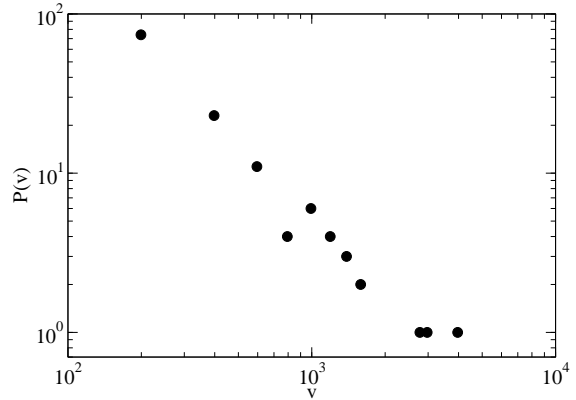,height=6.5cm,angle=0}  
}
\caption{Distribution of the total daily volume of transaction per bank. This quantity is used as fitness in our model}
\label{PdV}
\end{figure}

Following the Pareto's law (confirmed in this data analysis) we assume that the 
distribution of sizes $v$ in the model is a power-law $P(v)\propto v^{-2}$, where the 
value of the exponent correspond to that of the data (see Fig.\ref{PdV}).  

We assign to the $N$ nodes ($N$ is 
the size of the system) a value drawn from the previous distribution. Vertices origin and 
destination for one edge are chosen with a probability $p_{ij}$ proportional to the sum 
5B
of respective sizes $v_i$ and $v_j$. In formulas 
\begin{eqnarray}
p_{ij}&=&\frac{(v_i+v_j)}{\sum_{i,j>i} (v_i+v_j)}.\\ 
\sum_{i,j>i}(v_i+v_j)&=&\frac 1 2 \sum_{i,j\ne i}(v_i+v_j)=(N-1)V_{tot}\\
\mbox{where } V_{tot}& =&\sum_j v_j
\end{eqnarray}
We obtain in this way $p_{ij}=\frac{v_i+v_j}{(N-1)V_{tot}}$
This choice of probability reproduces the fact that big banks are privileged in 
transactions among 
themselves while two little banks are very unlikely to interact.  
We produce an ensemble of $100$   statistical realizations of the model and then we 
calculate average statistical distributions. 
In Fig. 2 we compare experimental and simulated $P(k)$, $c(k)$ and $k_{nn}(k)$: here 
the distributions are also averaged on all EOM days of 2002. 
The simulation of the model reproduces remarkably well the considered topological 
properties of the inter-bank market $P(k)$, $c(k)$ and $k_{nn}(k)$.  
The real and simulated networks disclose disassortative behavior: this phenomenon has already 
been observed in other systems and it has been called {\em rich club phenomenon}, 
referring to the fact that in many real networks hubs are often connected each 
other\cite{CFSV06}. Fitness models on the other hand are known to produce disassortative 
networks, even if with different fitness distributions\cite{SBC04}.

It is interesting to note that this model does not consider preferential attachment rules. 
With the term ``preferential attachment'' it is indicated a specific procedure in which a 
vertex receives more edges according to the value of its degree. 
Note that this procedure must be very precise because if the probability of growth is 
proportional to the degree raised to a power different from $1$, the scale invariance is 
destroyed. Therefore, preferential attachment has a precise definition different from 
``rough proportionality''. When considering instead a fitness algorithm, it is true 
that the largest the fitness the largest the degree, but the microscopic procedure is 
different. A large degree is a consequence of an intrinsic quality, not the cause of the 
improvement of site connectivity. This is an important point since in this way the search for 
the origin of scale-invariance in networks can be explained by means of the ubiquitous 
presence of Pareto's law in Economics and Finance.

To quantify the agreement between experimental and simulated networks we also 
define an overlap parameter $m$ specifying how good is the behavior 
of the model in reproducing the observed clustering.

To quantify the agreement between experimental and simulated networks, we proceed 
in the following way. 
We define a matrix $E$, that is a weighted matrix $4\times4$, where the weights represent 
the number of connections between groups. 
In order to measure the overlap between the matrices obtained by data and by computer model, 
we define a distance based on the differences between the elements of the matrices. 
\begin{equation}
d= \sum_{g,k \geq g} |E^{exp}_{g,k}-E^{num}_{g,k}|
\end{equation}
The sum of all elements, 
$\sum_{g,k \geq g} E^{exp}_{g,k}$ and $\sum_{g,k \geq g} E^{num}_{g,k}$, 
is equal to $E_{tot}$ in both cases. Therefore the maximum possible difference is 
$2E_{tot}$. This happens when all the links are between two groups in one case and in 
other two groups in the other. 
We use this maximum value to normalize the above expression and we than 
define the overlap parameter $m$:
$m = 1- d/2E_{tot}$

A natural way to define groups in the model is to obtain a similar number $c$ of banks for 
each class i.e. $c=N_{banks}/N_{classes}$. It is useful nevertheless to pass to
continuous form. Using the previously introduced $P(v)$ giving the probability
distribution of the size $v$ of one bank. Banks of the same group $g$ are in
the range $[v_g,v_g+\triangle v_g]$.
\begin{equation}
\int_{v_g}^{v_g+\triangle v_g}P(v')dv'=c
\end{equation}
In our case, since the average number of banks is $140$, we obtain  $c\simeq 35$.
Then $\triangle v=cv^2/(N-cv)$. 
We now compute the number $E_{g,k}$ of links going from one group of banks 
$g_g$ to another one $g_k$, for every possible pair of banks. 
\begin{equation}
E_{g,k}= \sum_{i,j} a_{i,j} \delta (g_g-g(i)) \delta (g_k-g(j))
\end{equation}
where $g(i)$ represent the group of bank $i$ and $a_{i,j} $ is the element of the adjacency matrix. In the continuous approximation, 
defining $E_{v'v''}$ the number of edges from vertices of fitness $v'$ to vertices of fitness $v''$, $E_{g,k}$ 
is given by
\begin{eqnarray}
E_{g,k}&=&\int_{v_g}^{v_g+\triangle v_g}\int_{v_k}^{v_k+\triangle v_k} E_{v'v''}dv'dv''= \nonumber \\ 
&=&(N/2) \int\int P(v') P(v'')p(v',v'') dv'dv'' 
\end{eqnarray}
where $N$ is the number of vertices, 
$p(v',v'')$ is the linking probability, $P(v)$ is the fitness distribution and the 
formula is obtained 
integrating the expression for the average degree\cite{SBC04} 
(the integration domains are the ranges of volumes of groups $g$ and $k$ respectively). 
To evaluate the relevance of division in classes, we have to compare the value of  
$E_{g,k}$ with the corresponding quantity $E_{g,k}^{null}$ for a network where there is 
not a division in classes (null hypothesis). 
The analytical expression for the null case is 
$E_{g,k}^{null}=E_{tot}/10$
where $10$ is the number of possible couplings between the $4$ groups. 
The comparison between the two 
networks evidences that in the real case emerges the division in groups: 
in Tab. 1 for each possible combination of groups is reported the value  $E_{g,k}/E_{tot}$. 
In the null case, each element of the same matrix  should be equal to $10$.
In our case the overlap $m$ is very good ($98\%$).

\begin{table}[h]%[htpb]
\begin{centering}
\begin{tabular}{|c|c|c|c|c|}
\hline
Group & $1$  &  $2$  &  $3$  &  $4$  \\
\hline
 $1$  & $0$  &  $6$  &  $4$  &  $8$  \\
 $2$  & $6$  &  $3$  &  $8$  &  $17$ \\
 $3$  & $4$  &  $8$  &  $5$  &  $27$ \\
 $4$  & $8$  &  $17$  &  $27$  &  $22$ \\
\hline
\end{tabular}
\caption{(Color on line)The number of daily interactions between the banks 
of different groups. Data have been averaged during one month.}
\label{table_cluster}
\end{centering}
\end{table}

In conclusion we  present here a network representation of a financial market that 
in a natural way allows to measure the presence of clustering. By means of a suitable 
chosen model of network formation we can also understand the mechanism driving the 
formation of such clusters.  
The agreement between the model and experimental results is remarkably good;
this seems to suggest that the network formation is not due to the growth
mechanism of preferential attachment. 
Since the effects of European Central Bank policies are under debate\cite{BIR04}, graph 
theory can help in understand the system behavior under change of external conditions.
 
GC acknowledges support from European Project DELIS


\begin{thebibliography}{99}
\bibitem{EMID}
Data are taken from the Italian electronic broker market MID 
(Market for Interbank Deposits) is run  by e-MID S.p.A.  
Societ\`a  Interbancaria per l'Automazione (SIA), Milan. 
The Italian electronic broker Market for Interbank Deposit 
(e-MID) covers the entire overnight deposit market in Italy. 
The information about the parties involved in a
transaction allows us to perform an accurate daily analysis of 
the connectivity among banks and its change over time.
\end{thebibliography}
\end{document}